\documentclass[preprint,12pt]{elsarticle}

\usepackage{graphicx,subfigure}
\usepackage{amssymb}
\usepackage{amsmath}
\usepackage{bm}
\usepackage{lineno} 
\usepackage{natbib}
\usepackage[squaren]{SIunits}
\journal{J. Comp. Phys.} 

\begin{document}

\begin{frontmatter}

\title{Lorentz boosted frame simulation of Laser wakefield acceleration in quasi-3D geometry}

\author[UCLAEE]{Peicheng Yu}
\ead{tpc02@ucla.edu}
\author[UCLAEE,UCLAPH]{Xinlu Xu} 
\author[UCLAPH]{Asher Davidson}
\author[UCLAPH]{Adam Tableman}
\author[UCLAPH]{Thamine Dalichaouch}
\author[UCLAPH]{Michael D. Meyers}
\author[UCLAPH]{Frank S. Tsung}
\author[UCLAPH]{Viktor K. Decyk}
\author[SLAC]{Frederico Fiuza}
\author[IST]{Jorge Vieira}
\author[IST,ISCTE]{Ricardo A. Fonseca}
\author[THUACC]{Wei Lu}
\author[IST]{Luis O. Silva}
\author[UCLAEE,UCLAPH]{Warren B. Mori}

\address[UCLAEE]{Department of Electrical Engineering, University of California Los Angeles, Los Angeles, CA 90095, USA}
\address[UCLAPH]{Department of Physics and Astronomy, University of California Los Angeles, Los Angeles, CA 90095, USA}
\address[SLAC]{SLAC National Accelerator Laboratory, Menlo Park, CA 94025, USA}
\address[IST]{GOLP/Instituto de Plasma e Fus\~ao Nuclear, Instituto Superior T\'ecnico, Universidade de Lisboa, 1049--001, Lisbon, Portugal}
\address[ISCTE]{ISCTE - Instituto Universit\'ario de Lisboa, 1649--026, Lisbon, Portugal}
\address[THUACC]{Department of Engineering Physics, Tsinghua University, Beijing 100084, China}

\begin{abstract}
When modeling laser wakefield acceleration (LWFA) using the particle-in-cell (PIC) algorithm in a Lorentz boosted frame, the plasma is drifting relativistically at $\beta_b c$ towards the laser, which can lead to a computational speedup of $\sim \gamma_b^2=(1-\beta_b^2)^{-1}$. Meanwhile, when LWFA is modeled in the quasi-3D geometry in which the electromagnetic fields and current are decomposed into a limited number of azimuthal harmonics, speedups are achieved by modeling three dimensional problems with the computation load on the order of two dimensional $r-z$ simulations. Here, we describe how to combine the speed ups from the Lorentz boosted frame and quasi-3D algorithms. The key to the combination is the use of a hybrid Yee-FFT solver in the quasi-3D geometry that can be used to effectively eliminate the Numerical Cerenkov Instability (NCI) that inevitably arises in a Lorentz boosted frame due to the unphysical coupling of Langmuir modes and EM modes of the relativistically drifting plasma in these simulations. In addition, based on the space-time distribution of the LWFA data in the lab and boosted frame, we propose to use a moving window to follow the drifting plasma to further reduce the computational load. We describe the details of how the NCI is eliminated for the quasi-3D geometry, the setups for simulations which combine the Lorentz boosted frame and quasi-3D geometry, the use of a moving window, and compare the results from these simulations against their corresponding lab frame cases. Good agreement is obtained, particularly when there is no self-trapping, which demonstrates it is possible to combine the Lorentz boosted frame and the quasi-3D algorithms when modeling LWFA to achieve unprecedented speedups. 
\end{abstract}


\end{frontmatter}

%% main text

%%%%%% %%%%%%%%%%%%%% INTRODUCTION %%%%%%%%%
\section{Introduction}
\label{sect:intro}
Laser wakefield acceleration (LWFA) \cite{LWFA} offers the potential to construct compact accelerators that have numerous potential applications, including the building blocks for a next generation linear collider and the electron beam source for ultra-compact XFELs. It has thus attracted extensive interest, and the last decade has seen an explosion of experimental results. Fully nonlinear particle-in-cell (PIC) simulations have been instrumental in this progress as an aid in designing new experiments, in interpreting experimental results, and in testing new ideas. Furthermore, developing predictive theoretical models is challenging due to the strong nonlinear effects that are present in the blowout regime of LWFA \cite{Luetal}; therefore numerical simulations are also critical in exploring the physics of LWFA. Particle-in-cell simulations have been extensively applied in LWFA research because the PIC algorithm follows the self-consistent interactions of particles through the electromagnetic (EM) fields directly calculated from the full set of Maxwell equations. When modeling LWFA using the PIC algorithm the laser wavelength needs to be resolved which is usually on the scale of 1 $\micro\meter$; meanwhile, the length of the plasma column that the laser propagates through can be on the scale of $10^4$ to $10^6$ $\micro\meter$. As a result, of this disparity in cell size and propagation distance, full three-dimensional (3D) PIC simulations of LWFA can be very CPU-time consuming. To capture the key physics while reducing the computation time, reduced models are continually being proposed. These include models that combine the ponderomotive guiding center with full PIC for the wake \cite{pgc} or with the quasi-static approach \cite{Mora,CK}. However, these models cannot as yet model full pump depletion lengths, and the quasi-static approach cannot model self-injection. 

Recently, two methods have been proposed that can speed up the LWFA simulation without losing key physics in the modeling of LWFA. One method is the Lorentz boosted frame technique \cite{Vay2007PRL}. In this method the LWFA simulations are performed in an optimized Lorentz boosted frame with velocity $v_b$, in which the length of the plasma column is Lorentz contracted, while the laser wavelength is Lorentz expanded. Assuming the reflection of the laser light is not important in the lab frame, then in a properly chosen Lorentz transformed frame the time and space scales to be resolved in a numerical simulation are minimized, and savings of factors of $\gamma^2_b=(1-v^2_b/c^2)^{-1}$ can be achieved. Another method that has been recently proposed is to expand the fields in to azimuthal harmonics and to truncate the expansion \cite{Lifschitz,davidson}. This can reduce the computational costs of modeling  3D problem with low azimuthal asymmetry to that on the order of 2D $r-z$ simulations. 

It was pointed out in Ref. \cite{YuAAC14,YuIPAC15} that it would be intriguing to combine these two methods in order to combine the speedups provided by each. Similarly to full PIC simulations in the Cartesian geometry, it was found that in the quasi-3D geometry one of the obstacles to performing Lorentz boosted frame simulations is the multi-dimensional Numerical Cerenkov Instability (NCI) \cite{Godfrey1974,YuAAC12,GodfreyJCP2013,XuCPC13} that inevitably arises due to the unphysical coupling between Langmuir modes (main and aliasing) and EM modes of the relativistic drifting plasma in the simulations. The coupling arises in the Lorentz boosted frame between modes which are purely longitudinal (Langmuir modes) and purely transverse (EM modes) in the lab frame. The coupling occurs at specific resonances $(\omega - \mu2\pi/\Delta_t)=(k - \nu_z 2 \pi/\Delta_z)v_b$ where $\mu$ and $\nu_z$ are the time and space aliases and $\Delta_t$ and $\Delta_z$ are the time step and grid size respectively. 

While the multi-dimensional NCI theory in Cartesian coordinates has been well studied (see e.g. \cite{YuAAC12,GodfreyJCP2013,XuCPC13,GodfreyJCP2014,GodfreyJCP20142,YuCPC15,YuHybrid15}), there are currently no analytical expressions for the numerical dispersion relation of relativistic plasma drifting in the quasi-3D geometry. 
However, OSIRIS \cite{osiris} simulations have shown that its behavior for the quasi-3D $r-z$ geometry is very similar to that in Cartesian geometry. It was therefore recently proposed and demonstrated that a hybrid Yee-FFT solver could be used to suppress the NCI in the Cartesian and quasi-3D geometries \cite{YuHybrid15}. In the regular Yee (a finite difference) solver in a quasi-3D geometry \cite{Lifschitz,Yee}, Maxwell equations are solved in $(r,z)$ space for each azimuthal mode $m$. In the hybrid Yee-FFT solver,  we perform a Fourier (discrete) transform in the drifting direction of the plasma (denote as $\hat z$), and solve Maxwell equations in $k_z$ space for each mode $m$; meanwhile, in the $\hat r$ direction the derivatives are represented as second order finite difference operators on a Yee grid. The current is corrected to maintain the correctness of Gauss' Law. When Maxwell's equations are solved in this way, the corresponding NCI modes can be systematically eliminated by applying the same strategies used for a multi-dimensional spectral Maxwell solver \cite{XuCPC13,YuCPC15}. The fastest growing modes of the NCI at $(\mu,\nu_1)=(0,\pm 1)$ can be conveniently suppressed by applying a low-pass filter in the current, the highly localized $(\mu,\nu_z)=(0,0)$ NCI modes can be moved away from physical modes by reducing the time step, and can be completely eliminated by modifying the EM dispersion at the $k_z$ range where the $(\mu,\nu_z)=(0,0)$ NCI modes are located. Furthermore, higher order spatial aliasing NCI modes can be suppressed by applying higher order particle shapes.  We present OSIRIS  simulation results which show that Lorentz boosted simulations of LWFA can be performed in this geometry with no evidence of NCI. It is worth noting that recently a PIC algorithm based on a fully spectral solver in quasi-3D geometry has been proposed by Lehe et. al. \cite{Lehe}. This scheme was demonstrated with a single-node algorithm.

In addition, according to how the lab frame information is located in the $(z',t')$ space, we show that the computation loads can be further reduced by applying a moving window in the boosted frame simulation. In the boosted frame the window follows the plasma as opposed to the laser, which is the case when using a moving window in the lab frame. 

The remainder of this paper is organized as follows: in section \ref{sect:hybridsolver} we briefly discuss the hybrid Yee-FFT solver in quasi-3D geometry, and the corresponding NCI mitigation strategies. In section \ref{sect:simsetup}, we discuss the simulation setups for modeling LWFA in the Lorentz boosted frame. We discuss the distribution of the data needed for the reconstruction of lab frame information with an emphasis on showing that using a moving window in the direction of the plasma drift can further reduce the computation load. We then show sample quasi-3D simulations of LWFA in the Lorentz boosted frame in section \ref{sect:sample}, and compare the results with the corresponding lab frame data. Good agreement is obtained demonstrating that the Lorentz boosted frame technique together with the quasi-3D algorithm and a moving window can be combined to achieve unprecedented speed up. The results are summarized in section \ref{sect:summary}.

\section{Hybrid Yee-FFT solver in quasi-3D geometry}
\label{sect:hybridsolver}
A key issue that needs to be addressed when performing LWFA simulations in a Lorentz boosted frame is the existence of a violent numerical instability, called the Numerical Cerenkov Instability (NCI). The NCI  arises when a plasma drifts relativistically on the grid. There has been much recent progress in identifying the NCI as the source of the instability, in deriving the numerical dispersion relations and determining growth rates, and in identifying mitigation strategies \cite{Godfrey1974,YuAAC12,GodfreyJCP2013,XuCPC13,GodfreyJCP2014,GodfreyJCP20142,YuCPC15,YuHybrid15}. In Ref. \cite{YuHybrid15} a hybrid Yee-FFT solver was proposed for the elimination of the NCI in Cartesian geometry. In this solver,  Maxwell equations are Fourier transformed in the drifting direction of the plasma (denoted as the $\hat z$ direction). The fields are solved in the corresponding $(k_z, x, y)$ space, where conventional second order finite difference operators on a Yee mesh are used in $(x,y)$. When Maxwell equations are solved in this way, the corresponding EM dispersion of the solver leads to NCI patterns that are very similar to those from a fully spectral Maxwell solver in which Maxwell equations are solved in multi-dimensional $\vec k$-space. Therefore one can systematically eliminate the NCI using approximately the same strategies developed for a fully spectral solver. Importantly, the hybrid Yee-FFT solver works for both Cartesian geometry $(z, x,y)$, and quasi-3D geometry $(z,r,\phi)$.

When the field solver is modified from a standard Yee solver to a hybrid Yee-FFT solver, essentially the EM dispersion in the $\hat z$ direction is modified from second-order accuracy (derived from its finite difference form) into a greater than $N$-th order accuracy. However, in OSIRIS (and most of the modern PIC codes) the $\vec E$ and $\vec B$ fields are advanced using Faraday's Law and Ampere's Law, while Gauss's Law is satisfied by applying a charge conserving current deposition scheme \cite{davidson,chargeconserving}. If the continuity equation is rigorously satisfied at each time step then by taking the finite difference version of Ampere's law, Gauss' Law is satisfied if it is true at $t=0$. However, the rigorous charge conserving current deposit is known only for second order finite difference operators. Therefore, when we use a FFT for the differential operator along $\hat z$ direction in Faraday's and Ampere's Law, we need to modify the current appropriately so the continuity equation is still true for the modified differential operator. To accomplish this, we Fourier transform the current for each azimuthal mode, $\vec J^m$, along $\hat z$-direction obtained from the current deposition scheme described in \cite{davidson}, and then apply a correction of the form,
\begin{align}
\tilde J^m_z = \frac{\sin k_z\Delta z/2}{k_z\Delta z/2}J^m_z
\end{align}
before we use it in the solver. This correction is applied to each azimuthal mode $m$ in the simulation to ensure that Gauss's Law is satisfied for each $m$ mode. 

We then Fourier transform $\vec E$ and $\vec B$ along $\hat z$-direction, and solve Faraday's Law and Ampere's Law for each azimuthal mode $m$, and each Fourier mode $k_z$, using the corrected current as the source term, 
\begin{align}
\partial_t B^m_r&=-\frac{im}{r}E^m_z-ik_z E^m_\phi\\
\partial_t B^m_\phi &=ik_z E^m_r+ \partial_r E^m_z\\
\partial_t B^m_z &=-\frac{1}{r} \partial_r(rE^m_\phi)+\frac{im}{r}E^m_r\\
\partial_t E^m_r &=\frac{im}{r}B^m_z+ik_z B^m_\phi-J^m_r\\
\partial_t E^m_\phi &=-ik_z  B^m_r-\partial_r B^m_z-J^m_\phi\\
\partial_t E^m_z&=-\frac{1}{r}\partial_r(rB^m_\phi)-\frac{im}{r}B^m_r-\tilde J^m_z
\end{align}
Note that $\partial_t$ and $\partial_r$ adopt the conventional finite difference form as in the Yee solver. The code is gridless in $\phi$ so $\partial_\phi$ is replaced with $im$.

We have found previously that the NCI pattern for the quasi-3D hybrid Yee-FFT solver is similar to its counterpart in the Cartesian 3D geometry \cite{YuAAC14,YuHybrid15}. As a result, we can apply approximately the same mitigation strategies used for the fully spectral solver in Cartesian geometry to systematically eliminate the NCI modes for this solver \cite{XuCPC13,YuCPC15}. We first eliminate the fastest growing $(\mu,\nu_z)=(0,\pm 1)$ modes ($\nu_z$ is the spatial aliasing in $\hat z$ direction) by applying a low-pass filter in the current. This can be easily accomplished since the current density is already in $k_z$ space after the Fourier transform. The second fastest growing NCI modes $(\mu,\nu_z)=(0,0)$ can be eliminated by reducing the simulation time step. Using a reduced time step not only reduces the $(\mu,\nu_z)=(0,0)$ NCI growth rate but also moves their locations away from the  modes where important physics is occurring. Note that this cannot be done using a pure finite difference solver since numerical dispersion errors for light waves get worse as the time step is reduced if the cell sizes remain fixed. The $(\mu,\nu_z)=(0,0)$ NCI modes can also be eliminated by slightly modifying the $k_z$ operator to create a small bump in the dispersion relation to precisely avoid intersections between the main EM modes and main Langmuir modes at that highly localized region in $k_z$. As for higher order NCI modes, their growth rates can be reduced by applying higher order particle shapes. Applying the strategies described above, we have found that we can systematically eliminate the NCI modes in the quasi-3D geometry.

Fourier transforming the current into $k_z$ space is not only important for the efficient filtering of the NCI modes, but is also required to accurately correct (compensate) the current in $k_z$ space to exactly match the modified Maxwell solver. It is worth noting that it is now a common practice to modify either the Maxwell solver or the field interpolation to change the EM dispersion relation in order to obtain a more desirable dispersion relation \cite{YuAAC12,GodfreyJCP2013,XuCPC13,GodfreyJCP2014,GodfreyJCP20142,YuCPC15,YuHybrid15}. Within these schemes, Gauss' Law is satisfied by either directly solving it  (as is the case in UPIC \cite{dawsonrmp,lin,upic}), or by using a current that satisfies the continuity equation through a correction (compensation) to match the current deposition scheme with the Maxwell solver (as is the case in here and in \cite{YuHybrid15}). In the approaches which aim to eliminate the NCI \cite{YuHybrid15,YuJCP14} (rather than reducing its growth rate for any time step size to that for an optimum time step \cite{GodfreyJCP2014}), the use of an FFT is inevitable (at least in one direction) since it is very challenging to design a current deposition scheme in real space to match the EM dispersion of a modified Maxwell solver for an arbitrary order finite difference scheme.

\section{Simulation setups in the boosted frame}
\label{sect:simsetup}
The setup of a  quasi-3D LWFA simulation in a Lorentz boosted frame is almost identical to its counterpart in Cartesian 2D/3D geometry. In a boosted frame with Lorentz factor $\gamma$ that moves in the propagation direction of the laser, the laser pulse is colliding with a counter-propagating relativistically drifting plasma \cite{Martins2010NatPhys,VayJCP2011,MartinsCPC}. Due to the Lorentz transform, the plasma density increases by $\gamma$ while the total plasma column length contracts by $\gamma$. The laser wavelength and pulse length stretch by $\gamma(1+\beta)$, while its Rayleigh length contracts by $\gamma$. To avoid initializing a laser with very wide transverse size due to the contracted Rayleigh length and stretched pulse length, a moving antenna is placed at the edge of the plasma boundary to inject a laser pulse into the plasma \cite{VayPoP20112,VayAAC2010}. 

\subsection{Relationship between lab and boosted frame data}
\begin{figure}[t]
\begin{center}\includegraphics[width=1\textwidth]{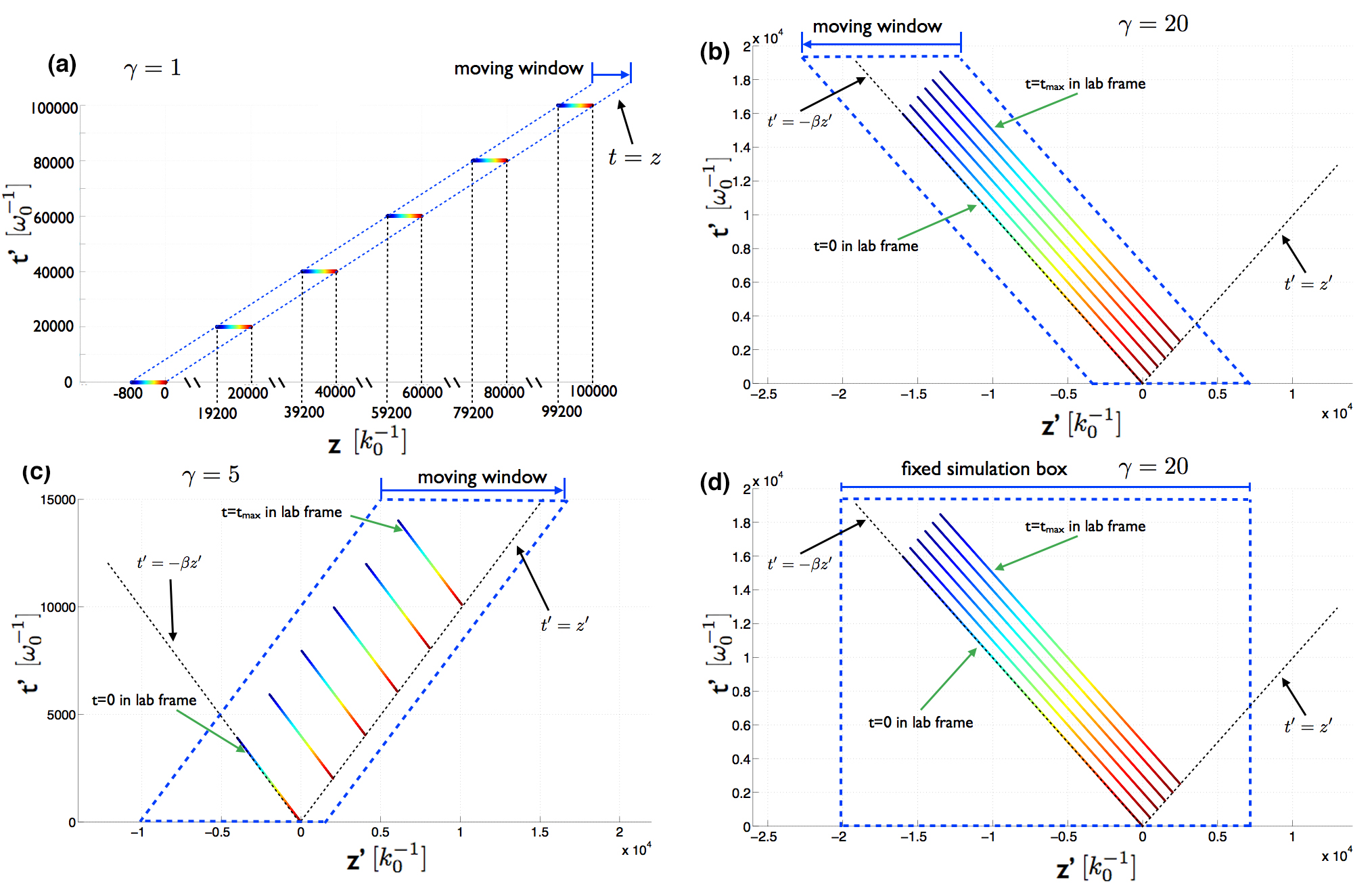}
\caption{\label{fig:x1trange} Range of important data in lab and boosted frame simulations. (a) Range of data in lab frame (stationary plasma) simulation with and without a moving window, (b) range of data in a boosted frame simulation with $\gamma_b=20$ including with a moving window, (c) range of data in a boosted frame simulation with $\gamma_b=5$ with a moving window, and (d) range of data in a boosted frame simulation with $\gamma_b=20$ without a moving window.} 
\end{center}
\end{figure}

In  LWFA simulations in the lab frame (i.e., a stationary plasma) the use of a moving window \cite{movingwindow}, which only follows the physical domain near the laser, significantly reduces the computational load. The moving window essentially drops plasma sufficiently far behind the laser and adds fresh plasma in front of the laser.  This is illustrated in Fig. \ref{fig:x1trange} (a) where we plot the range of space time data from a lab frame simulation. The solid box shows the total space time area while the dashed box shows the reduced area through the use of a moving window.  The moving window has a length $800~[k^{-1}_0]$, and the simulation duration is $t_{\max}=100000~[\omega^{-1}_0]$. We also show the simulation data that is dumped as colored lines. The data is dumped every $20000~[\omega^{-1}_0]$. The red ends of the data lines indicate the starting end of the moving window, while the blue ends indicate the rear end. Connecting the red ends of the data lines, we obtain the $z-t$ relation for the head of the moving windows, $t=z$ (the speed of light $c$ is normalized to 1). The data obtained in the lab frame (assuming the code dumps data at a constant time interval) rotates in space-time in the boosted frame since the Lorentz transform is essentially a hyperbolic rotation of coordinates in Minkowski space \cite{MartinsCPC,VayPoP2011}. Therefore lines of data in $\hat z$ taken at fixed time from a Lorentz boosted frame are rotated by the Lorentz transform, i.e, $t'=t/\gamma_b-\beta_b z'$. The slope of each data line now becomes $-\beta_b$, where $\beta_b=(1-\gamma_b^{-2})^{-1/2}$ and each data line in the lab frame which belongs to the same point in time in lab frame is now spread over a range of $t'$ and $z'$. Interestingly, when we connect the red end of each data line in the boosted frame it still has a slope of $c$, i.e. $t'=z'$. The range of data in the boosted frame is shown in Figs. \ref{fig:x1trange} (b), (c) and (d). The data  in Fig. \ref{fig:x1trange} (b), (c) corresponds to $\gamma_b=20$ while that in  Fig. \ref{fig:x1trange} (d) corresponds to $\gamma_b=5$. In Figs. \ref{fig:x1trange} (b) and (c) we also show the smallest area (domain enclosed by dashed lines) in $t',z'$ space that includes the area needed to reconstruct the lab frame data for the two different values of $\gamma_b$. This illustrates that the space-time area in the boosted frame can be minimized by using a moving window in this frame. In Fig. \ref{fig:x1trange} (b) it is seen that this window moves to the left (backwards); while in Fig. \ref{fig:x1trange} (c) the window moves to the right (forwards). We use such moving window in the boosted frame OSIRIS simulations. Currently,  UPIC-EMMA boosted frame simulations in Cartesian 2D/3D geometry uses a stationary window [Fig. \ref{fig:x1trange} (d)] \cite{YuJCP14}. 

From Fig. \ref{fig:x1trange} it is evident that in lab frame simulations we usually dump data sparsely in time (large time intervals between time outputs), but the data at each grid is dumped at each time output. On the other hand, in order to recover the equivalent lab frame data in a boosted frame simulation, we need to sample boosted frame data at a much higher rate in time, but only need a small number of spatial locations. This can be seen by plotting a line across $z'$ for a fixed $t'$. This line only intersects the equivalent lab frame data at the same number of spatial locations as the number of time outputs. We typically dump the boosted frame data in a standard form (all grid points at small number of time steps) as well as the data needed to transform the results back to the lab frame (a small number of interpolated grid points at a large number of time steps). We then post-process the later data by performing the inverse rotation  back into lab frame for comparison with the lab frame data. When running in the lab frame we also plot the necessary data needed to reconstruct the data into a boosted frame. This inverse construction method is useful during the development of a boosted frame code, as one can transform the lab frame data that has been extensively cross checked with theory, to the boosted frame, and compare the results against the results obtained by the boosted frame code.

\subsection{Basic setup}
\begin{figure}[th]
\begin{center}\includegraphics[width=1\textwidth]{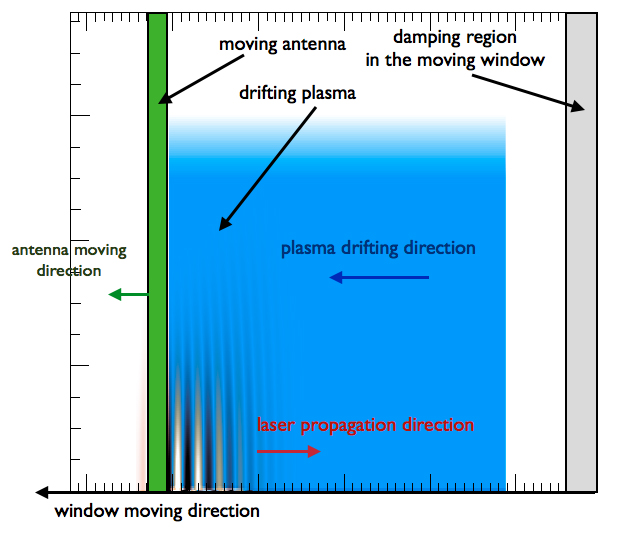}
\caption{\label{fig:simsetup} Simulation setup for a typical LWFA simulation in the boosted frame. The moving window follows the drifting plasma moving from right to left. A moving antenna injects laser pulse from left to right, and a damping region is located at the rear end of the moving window which also moves from left to right.} 
\end{center}
\end{figure}

In Fig. \ref{fig:simsetup}, we present a typical setup for a boosted frame simulation. The moving window moves from right to left following the drifting plasma. The moving antenna is also moving from right to left and injects the laser pulse from the left plasma boundary into the plasma. We place a damping section at the rear (right) end of the moving window (there is a gap between the plasma and the damping region) to damp the EM field to zero in this region. This is done because periodic boundary conditions are applied in the $\hat z$ direction when using the hybrid Yee-FFT solver, which requires that the EM fields need to be zero at the rear end of the simulation window to match the fields at the opposite side; otherwise the EM field at the rear end will reappear at the starting end. We note that there will be a low level of EM reflection from the damping section. In an FFT solver, the group velocity of light in vacuum is greater than the speed of light, however, since the simulation window is moving at the speed of light and the drifting plasma is drifting ultra-relativistically away from it, the reflected energy is not able to catch up with the drifting plasma. Hence the physics inside the plasma will not be affected by the reflecting EM waves. We have compared cases with the moving window plus the damping regions against cases without moving window to confirm that the moving window plus damping region works \cite{YuIPAC15}. We also note that for high $\gamma_b$ boosted frame simulations, we find that the modified pusher described in Ref. \cite{VayPusher} is required in order to get the evolution of the bubble correct. As pointed out in Ref. \cite{VayPusher} the usual leap frog staggering leads to issues for the Lorentz force when there is near cancellation of the electric and magnetic forces for relativistically moving particles. Determining at what $\gamma_b$ the modified pusher in \cite{VayPusher} is needed is an area of future work.

\section{Sample simulations}
\label{sect:sample}

\begin{table}[t]
\centering
\begin{tabular}{lr}
\hline\hline
Plasma &\\
\quad density $n_p$& $8.62\times 10^{-4} n_0\gamma_b$\\
\quad length $L$ & $8.0\times 10^4 k^{-1}_0/\gamma_b$\\
Laser & \\
\quad pulse length $\tau$ & $ 86.9 k^{-1}_0\gamma_b(1+\beta_b)$\\
\quad pulse waist $W$ & $153.0 k^{-1}_0$\\
\quad polarization & circular\\
\quad normalized vector potential $a_0$ & 4.0\\
Lab frame simulation $(\gamma_b=1)$&\\
\quad grid size $(\Delta x_{1},\Delta x_{2,3})$ & $(0.2k^{-1}_0, 3.4k^{-1}_0)$\\
\quad time step $\Delta t/\Delta x_1$ & $0.995$\\
\quad number of grid (moving window) & $4000\times 512\times 512$ \\
\quad particle shape & quadratic\\
Boosted frame simulation $(\gamma_b=15.0)$&\\
\quad grid size $\Delta x_{z,r}$ & $0.1k^{-1}_0\gamma_b(1+\beta_b)$\\
\quad time step $\Delta t/\Delta x_{z}$ &0.125\\
\quad number of grid (moving window) & 2048$\times$512\\
\quad particle shape & quadratic\\
\hline\hline
\end{tabular}
\caption{Parameters for the 3D and quasi-3D LWFA  simulations in the Lorentz boosted frame (discussed in section \ref{sect:boostedframedata}). The laser frequency $\omega_0$ and  number $k_0$ in the lab frame are used to normalize simulation parameters. The density is normalized to the critical density in the lab frame, $n_0=m_e\omega^2_0/(4\pi e^2)$. Normalized vector potential $a_0$ for the laser has been converted to linear polarization.}
\label{tab:lwfapara0}
\end{table}

\begin{table}[t]
\centering
\begin{tabular}{lr}
\hline\hline
Plasma &\\
\quad density $n_p$& $1.433\times 10^{-4} n_0\gamma_b$\\
\quad length $L$ & $1.63\times 10^6k^{-1}_0/\gamma_b$\\
Laser & \\
\quad pulse length $\tau$ & $ 296.4k^{-1}_0\gamma_b(1+\beta_b)$\\
\quad pulse waist $W$ & $351.9k^{-1}_0$\\
\quad polarization & circular\\
\quad normalized vector potential $a_0$ & 4.44/3.0\\
Lab frame simulation $(\gamma_b=1)$&\\
\quad grid size $(\Delta x_{z},\Delta x_{r})$ & $(0.2k^{-1}_0, 4.74k^{-1}_0)$\\
\quad time step $\Delta t/\Delta x_z$ & $0.9974$\\
\quad number of grid (moving window) & $7920\times 1248$ \\
\quad particle shape & quadratic\\
Boosted frame simulation $(\gamma_b=26.88)$&\\
\quad grid size $\Delta x_{z,r}$ & $0.2k^{-1}_0\gamma_b(1+\beta_b)$\\
\quad time step $\Delta t/\Delta x_z$ &0.25\\
\quad number of grid (moving window) & 8192$\times$792\\
\quad particle shape & quadratic\\
\hline\hline
\end{tabular}
\caption{Parameters for the quasi-3D LWFA  simulations in the lab frame and Lorentz boosted frame (discussed in section \ref{sect:labframedata}). The laser frequency $\omega_0$ and  number $k_0$ in the lab frame are used to normalize simulation parameters. The density is normalized to the critical density in the lab frame, $n_0=m_e\omega^2_0/(4\pi e^2)$. Normalized vector potential $a_0$ for the laser has been converted to linear polarization.}
\label{tab:lwfapara1}
\end{table}

\begin{figure}[th]
\begin{center}\includegraphics[width=1\textwidth]{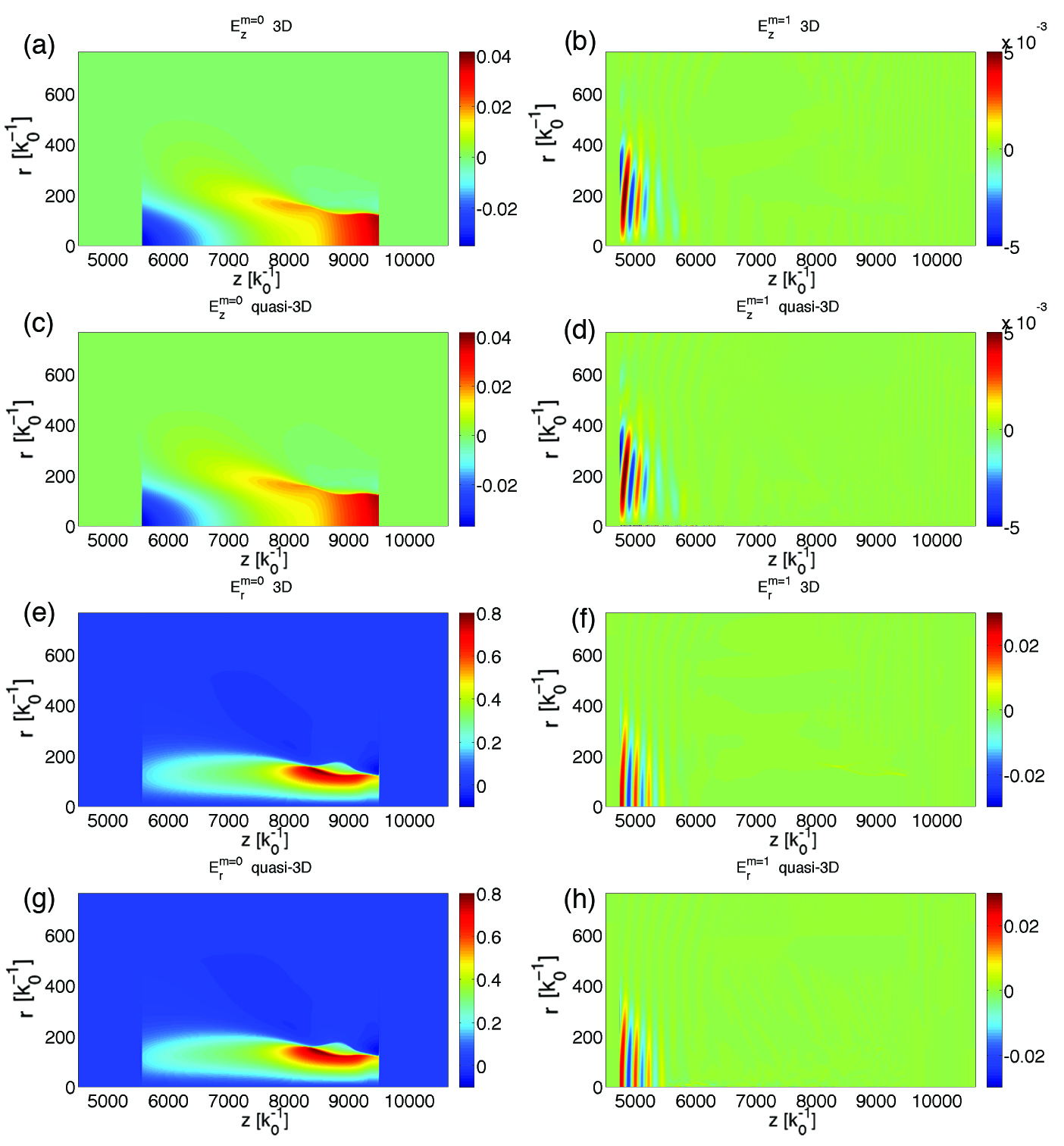}
\caption{\label{fig:3dvsq3} Comparison of simulation results in 3D and quasi-3D geometries for the $a_0=4.44$ (converted to linear polarization) 10 GeV LWFA stage run. All results are from boosted frame simulations. On the left are the $m=0$ modes from $E_z$ and $E_r$. On the right are the $m=1$ modes for $E_z$ and $E_r$. Results from a full 3D OSIRIS case are compared against a quasi-3D OSIRIS case where only $\vert m\vert\le 1$ modes were kept. } 
\end{center}
\end{figure} 

\begin{figure}[th]
\begin{center}\includegraphics[width=1\textwidth]{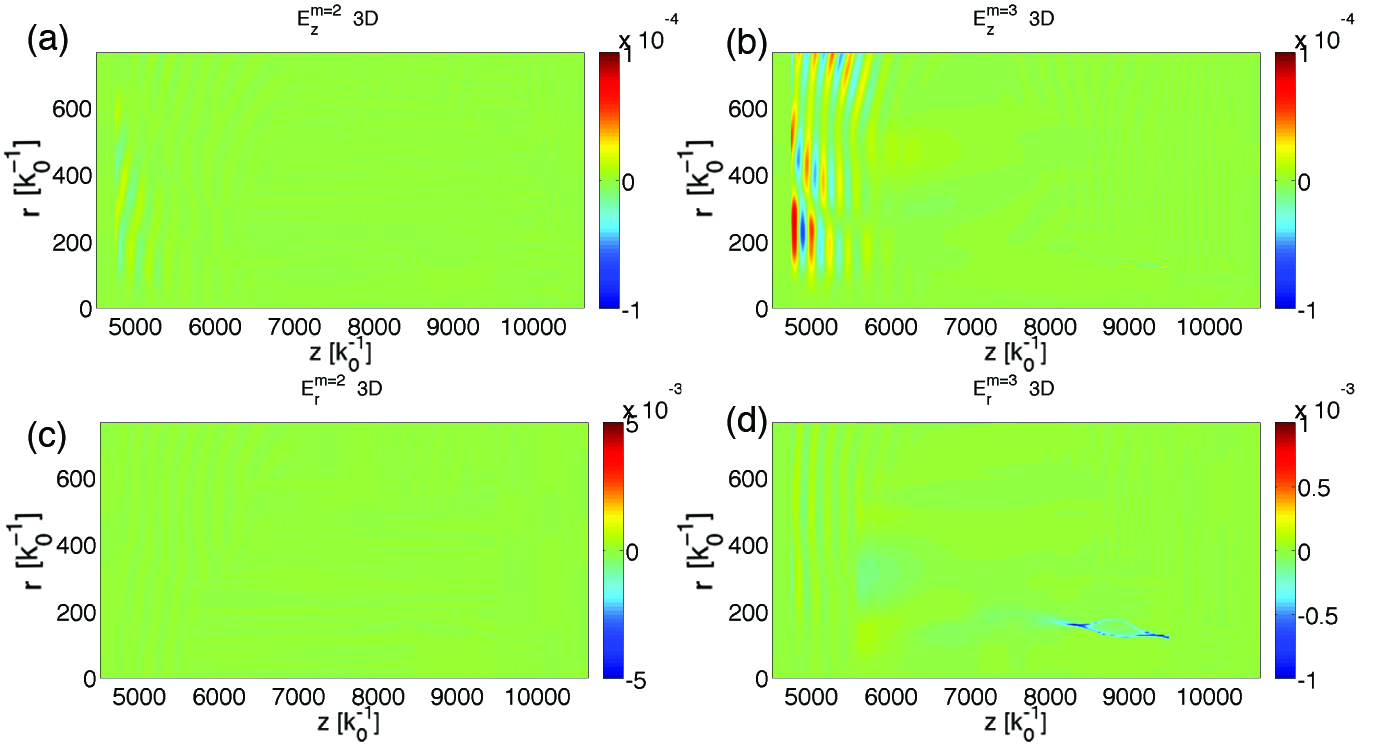}
\caption{\label{fig:higherorder} Higher order $m$ modes for $E_z$ and $E_r$ obtained from a full 3D LWFA boosted frame data. On the left are $E_z$ and $E_r$ for mode $m=2$, while of the right are $E_z$ and $E_r$ for mode $m=3$.} 
\end{center}
\end{figure} 

\begin{figure}[th]
\begin{center}\includegraphics[width=1\textwidth]{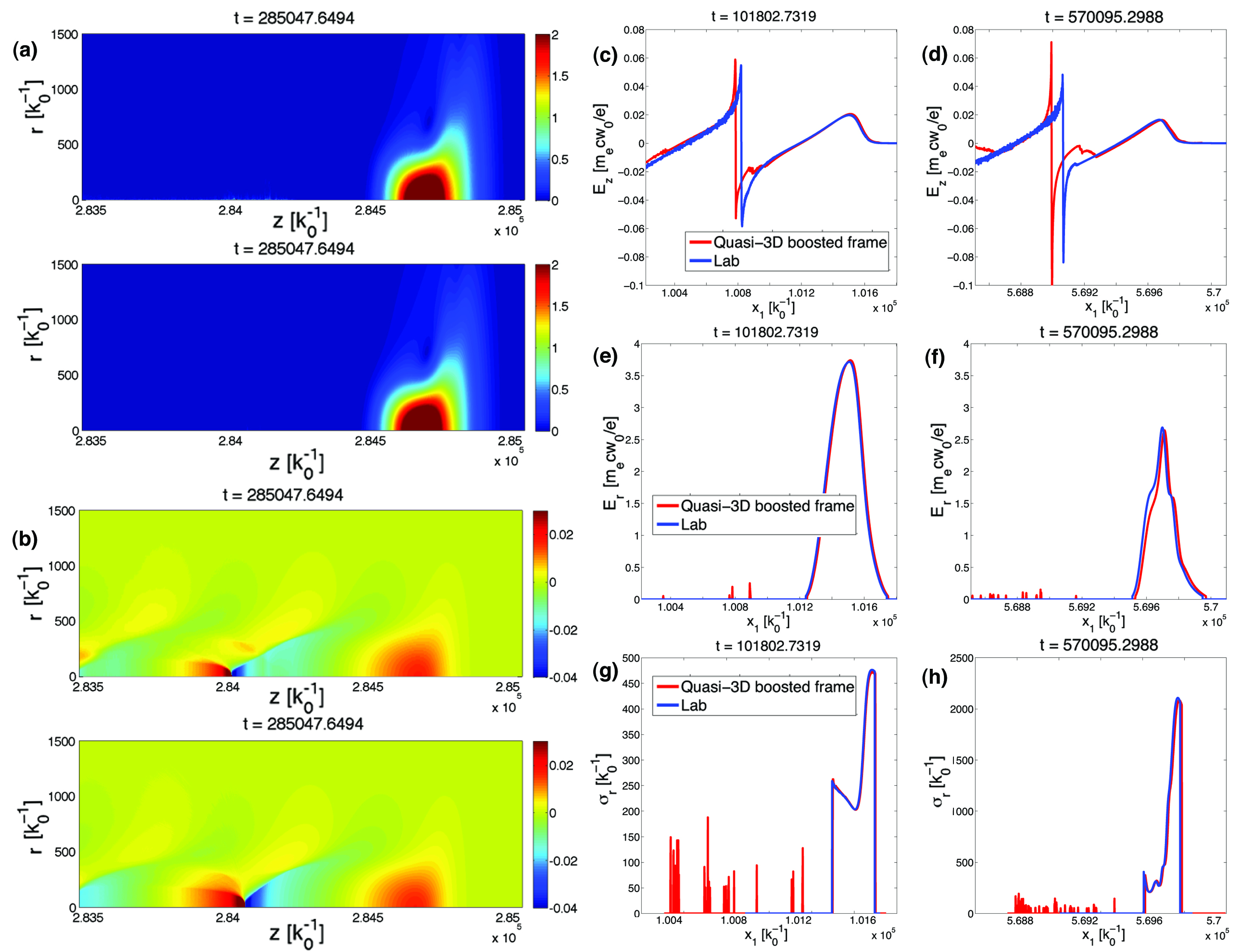}
\caption{\label{fig:a4} Simulation results for a $a_0=4.44$ (converted to linear polarization) 10 GeV LWFA stage run. (a) shows the comparison of envelope of $\mathcal{R}(E^{m=1}_r)$ field, which shows the evolution of laser driver as it propagates through the plasma; (b) shows the corresponding comparison of the amplitude of $\mathcal{R}(E^{m=0}_z)$, which shows how the wakefield of the bubble varies in the two frames due to the different self-injection results; (c), (e), and (g) are comparisons of line out for the wakefield, laser envelope, and laser spot size respectively at lab frame time $t=101802.7~\omega^{-1}_0$, while (d), (f), (h) are the corresponding plots at $t=570095.3~\omega^{-1}_0$. } 
\end{center}
\end{figure} 

\begin{figure}[th]
\begin{center}\includegraphics[width=1\textwidth]{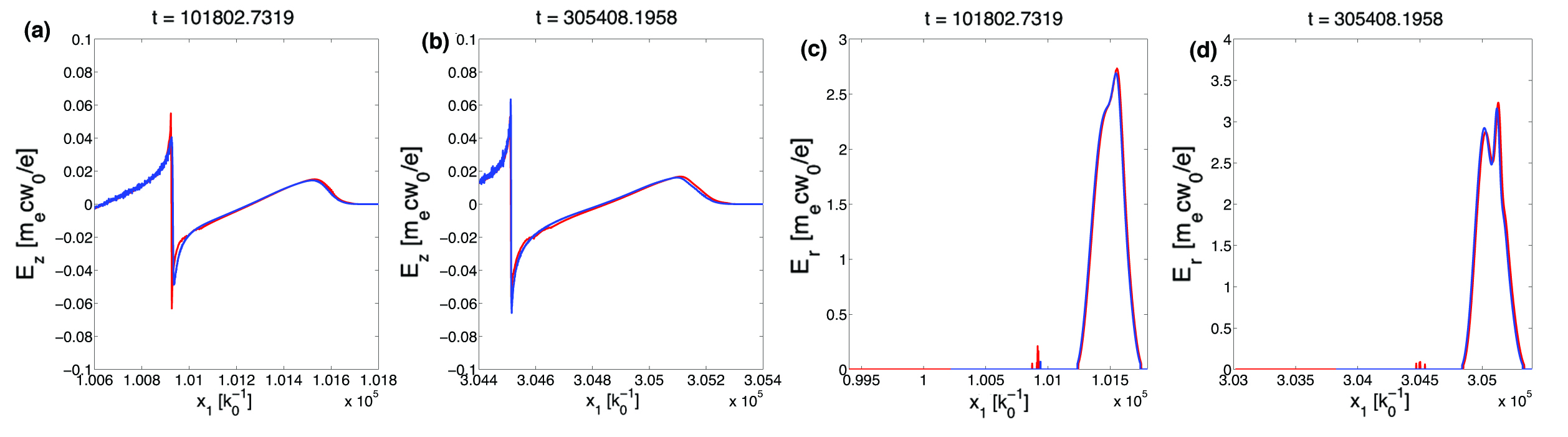}
\caption{\label{fig:a3} Line outs of wakefield and laser pulse shape at various lab frame time for a $a_0=3.0$ case is shown. Since there are no self-injection in the lab frame for this case, much better agreements are obtained for the wake field part.} 
\end{center}
\end{figure} 

In this section, we present two sample simulations. We begin by comparing results from two boosted frame simulations where in one case we use full 3D OSIRIS and in the second case we use quasi-3D OSIRIS. The parameters match those in Ref. \cite{Luetal} whereby a 200 TW laser is focused to a spot size of $19.5~\micro\meter$ at the entrance of a $1.5 \times 10^{18} ~\centi\meter^{-3}$ density plasma. The FWHM pulse length of the laser was 35 fs and the normalized vector was $a_0=4.0$ for a linearly polarized laser or $a_0=4.0/\sqrt{2}$ for a circularly polarized laser.  This corresponds to a 1.3 GeV output electron energy according to the scaling laws in Ref. \cite{Luetal}. The numerical parameters are shown in Table \ref{tab:lwfapara0}. We then compare the output in the boosted frame for various azimuthal mode numbers. This comparison requires the use of a post-processing algorithm which decomposes the full 3D data into azimuthal modes \cite{Thamine}. 

We then compare the data of a LWFA boosted frame simulation in quasi-3D lab with the corresponding quasi-3D boosted frame simulation. For this simulation we explore parameters for which a full 3D lab frame simulation is not feasible due to the large CPU hours required. The parameters correspond to a 1.8 PW laser  focused to a spot size of $45~\micro\meter$ at the entrance of a $2.5\times 10^{17}~\centi\meter^{-3}$ density plasma. The FWHM pulse length of the laser was $130~\femto\second$ and the normalized vector was $a_0=4.44$ for a linearly polarized laser or $a_0=4.44/\sqrt{2}$ for a circularly polarized laser. This corresponds to a 10.4 GeV output electron energy according to the scaling laws in Ref. \cite{Luetal}. The numerical parameters are shown in Table \ref{tab:lwfapara1}. The data from the boosted frame simulation is transformed back to the lab frame and it is compared against the data from the lab frame simulation.

\subsection{Comparison of data in the boosted frame}
\label{sect:boostedframedata}
When modeling LWFA in quasi-3D geometry, whether it is in the lab frame or boosted frame, the accelerating ($E_z$) and focusing fields ($E_r$ and $B_{\phi}$) in the bubble are mainly in the $m=0$ modes of the EM fields. On the other hand, the fields associated with the laser are associated with the $m=1$ mode of fields. Therefore, by keeping at least the $\vert m\vert \le 1$ modes the self-consistent evolution of the laser and wake fields can be examined when there is nearly azimuthal symmetry. For this comparison we truncate the azimuthal harmonics keeping only the $\vert m\vert \le 1$ modes \cite{Lifschitz,davidson}. More modes can be kept in principle to study laser hosing and asymmetric spot size effects as well as to test the convergence of the results. In addition, the results and the needed truncation can be verified by comparing LWFA boosted frame simulation results from the full 3D and quasi-3D geometries. To verify the azimuthal mode truncation, we decompose the data from a full 3D OSIRIS simulation into azimuthal harmonics and compare it against the corresponding quasi-3D simulation using the parameters listed in Table \ref{tab:lwfapara0}. In Fig. \ref{fig:3dvsq3}, we plot the azimuthal decomposition of the 3D data for $E_z$ and $E_r$ at $t'=4494.99~[\omega^{-1}_0]$, and compare it against the corresponding quasi-3D data at the same time. For the $\vert m\vert \le 1$ modes, very good agreement is observed. In addition, we plot the higher order $m=2,3$ modes from the 3D data in Fig. \ref{fig:higherorder}. We can see that the higher order modes are at least one order of magnitude smaller than those of the $m=0,1$ modes, which verifies the truncation of azimuthal harmonics at $\vert m\vert \le 1$ in the quasi-3D simulations when the laser is nearly symmetric. 

\subsection{Comparison of data in the lab frame}
\label{sect:labframedata}
Next, we compare data from a quasi-3D LWFA simulation in the lab frame against data Lorentz transformed back to the lab frame from a quasi-3D simulation. A laser with normalized vector potential of $a_0=4.44$ (converted to linear polarization) with pulse length of 130 fs, and spot size of 45 $\micro\meter$ propagates into a plasma column 20.8 cm long (in the lab frame). We use a boosted frame with $\gamma_b=26.88$, and use a moving window as described earlier that follows the relativistically drifting plasma. A moving antenna injects the laser pulse into the plasma, and a damping region absorbs the EM field at the rear end of the moving window. In the upper $\hat r$ boundary of the simulation box we applied the Perfectly-Matched-Layer boundary condition (see Ref. \cite{YuIPAC15} for more details). The plasma density is uniform along the $\hat z$ direction. It is uniform in $\hat r$ direction from $0\le r \le 7000~[k^{-1}_0]$ (where $k_0$ is the wave number of the laser in the lab frame), and then the density linearly ramps to zero at $r=8000~[k^{-1}_0]$ near the $\hat r$ upper boundary (an additional gap of $500~[k^{-1}_0]$ is left between the $\hat r$ upper plasma boundary and simulation box boundary). The linear plasma density ramp is used to prevent reflection when the laser cross the upper $\hat r$ plasma boundary into vacuum. Detailed simulation parameters are listed in Table \ref{tab:lwfapara1}.

As mentioned in section \ref{sect:simsetup}, in the boosted frame each azimuthal mode of the EM field is dumped frequently in time, while sparsely in space. The results are transformed back to the lab frame in the post-processing. In Figs. \ref{fig:a4} (a) and (b) we present 2D plots of the $m=0$ mode of the $E_z$ field, as well as the $m=1$ mode of the $E_r$ fields, while in Figs. \ref{fig:a4} (c) and (d) we present the corresponding data from the lab frame simulation, respectively. As we can see from Fig. \ref{fig:a4} (a) and (c) the data from the two simulations agree well with each other, except for the area around the rear of the first bubble, which indicates that the two simulations give different self-injection results. On the other hand, the laser profiles from the two cases agree extremely well. In Fig. \ref{fig:a4} (e) line outs of the $m=0$ mode for $E_z$ at various time steps are plotted, and they show that in the transformed boosted frame data there is stronger beam loading, which indicates that more charge is self-injected into the bubble. This is likely due to the difference in statistics between the lab frame simulation and boosted frame simulation. In the boosted frame a macro-particle represents much more charge than in the lab frame. Particles in the boosted frame are ``fatter'' since the grid size in the boosted frame is larger, and this could affect the self-injection process. To confirm the differences are related to the self-injection process, we repeated the lab frame and boosted frame simulations in regimes with no self-injection, at $a_0=3.0$ (converted to linear polarization), while keeping the other parameters listed in Table \ref{tab:lwfapara1} the same. In Fig. \ref{fig:a3} (a) and (b) we show the line out of the wakefield at two different times in the lab frame, and in Figs. \ref{fig:a3} (c) and (d) we show the corresponding amplitude of the laser profiles. We see from Fig. \ref{fig:a3} that for this case where there is no self-injection in the lab frame simulation, the wake field results from the lab frame and boosted frame simulations agree very well. It is challenging to accurately modeling the self-injection process in the LWFA blowout regime, and this is an area of future work. 

\begin{figure}[th]
\begin{center}\includegraphics[width=1\textwidth]{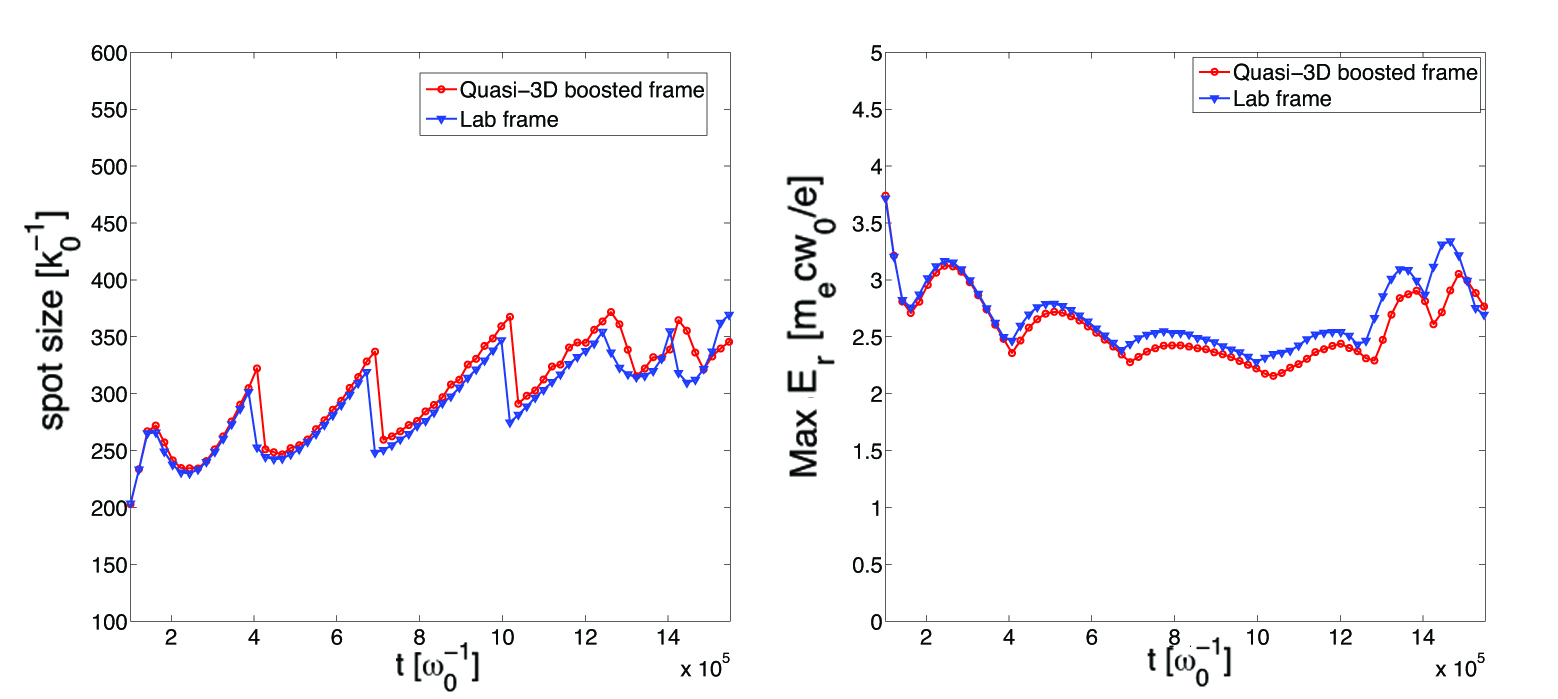}
\caption{\label{fig:spotamp} Evolution of the laser spot size and peak amplitude. (a) shows the comparison of laser spot size evolution as the laser propagates into the plasma for the two frames. The laser spot size are defined at the location where the laser has the maximum amplitude. The corresponding maximum laser amplitude evolution is shown in (b).} 
\end{center}
\end{figure}

As for the evolution of the laser profile, we plot the laser envelope and spot size obtained from the two cases from Figs. \ref{fig:a4} (e) and (f). Excellent agreement can be seen for the two times presented in Fig. \ref{fig:a4}. Excellent agreement is also seen for the evolution of the spot size, and laser amplitude of the laser driver as it propagates through the plasma column. In Fig. \ref{fig:spotamp} we show a detailed time history of the laser spot size and amplitude at the position of the laser where its amplitude is largest. Fig. \ref{fig:spotamp} clearly shows that very accurate results can be obtained when using Lorentz boosted frame technique in quasi-3D geometry to study the evolution of laser driver in the plasma, with a large savings in CPU time.

\section{Summary}
\label{sect:summary}
In this paper, we described how it is possible to perform LWFA simulations in Lorentz boosted frame using the quasi-3D algorithm. The key to high fidelity Lorentz boosted frame simulations in this geometry is the use of a hybrid Yee-FFT solver that solves the Maxwell equation in $k_z$ space in the direction that the plasma drifts, while keeping the second order finite-difference operators in the transverse directions as in a conventional Yee solver. Using this Maxwell solver we can then use the same strategies for eliminating NCI in Cartesian geometry to systematically eliminate the NCI in the quasi-3D geometry. At the same time all other features of OSIRIS are also available including single core optimization and high parallel scalability. A current correction is applied to ensure the code rigorously conserves charge. In addition, we analyzed the space-time area of the lab and boosted frame simulation data. We showed how using a moving window which follows the drifting plasma in the boosted frame the further reduce the computational load. We were able to combine Lorentz boosted frame technique with quasi-3D algorithm, together with moving window technique to achieve unprecedented speed up for the modeling of LWFA.

We presented comparisons of lab frame against boosted frame simulation results for a 10 GeV LWFA example that operates in the blowout regime. It was shown that the evolution of the laser driver in the plasma can be very well reproduced by the boosted frame simulation. We also found that the self-injection process is different in the boosted frame. This is partly due to the difference in statistics between the simulations in the two frames since in the boosted frame each macro-particle represent many more real particles then in the corresponding lab frame simulation. We found excellent agreement between the lab and boosted frame results for the wake fields when $a_0$ was reduced to avoid self-injection. An area of future work is to systematically explore methods to accurately model self-injection process in the Lorentz boosted frame simulation. Another area is the integration of this algorithm into our GPU and Intel-Phi enabled version of OSIRIS.

\section*{Acknowledgments}
This work was supported by US DOE under grants DE-SC0008491, DE-SC0008316, DE-FG02-92ER40727, and by the US National Science Foundation under the grant ACI 1339893, OCI 1036224, and by NSFC 11425521, 11535006, 11175102, 11375006, and Tsinghua University Initiative Scientific Research Program, and by the European Research Council (ERC-2010-AdG Grant 267841). Simulations were carried out on the UCLA Hoffman2 and Dawson2 Clusters, and on Hopper Cluster of the National Energy Research Scientific Computing Center, and on Blue Waters cluster at National Center for Supercomputing Applications at UIUC.

%%%%%%%%%%%%%%%%%%% bib %%%%%%%%%

\end{document}